\documentclass[aps,prl,twocolumn]{revtex4-1}
\usepackage{amsfonts,amssymb,amsmath,bm}
\usepackage{graphicx,epsfig,color}

\begin{document}

\title{Large-scale confinement and small-scale clustering of floating
particles in stratified turbulence}
\author{A. Sozza$^1$, F. De Lillo$^1$, S. Musacchio$^2$ and 
G. Boffetta$^1$}

\affiliation{
$^1$Department of Physics and INFN, Universit\`a di Torino,
via P. Giuria 1, 10125 Torino, Italy \\
$^2$Universit\'e de Nice Sophia Antipolis, CNRS, LJAD, UMR 7351, 
06100 Nice, France
}

\date{\today}

\begin{abstract}
We study the motion of small inertial particles in stratified turbulence. 
We derive a simplified model, valid within the Boussinesq approximation, 
for the dynamics of small particles in presence of a mean linear density 
profile. 
By means of extensive direct numerical simulations, we investigate
the statistical distribution of particles as a function of the two
dimensionless parameters of the problem.
We find that vertical confinement of particles is mainly ruled by
the degree of stratification, with a weak dependency on the particle
properties. Conversely, small scale fractal clustering, typical of
inertial particles in turbulence, depends on the particle relaxation
time and is almost independent on the flow stratification.
The implications of our findings for the formation of thin phytoplankton 
layers are discussed.
\end{abstract}

\pacs{}

\maketitle

%%%%%%%%%%%%%%%%%%%%%%%%%%%%%%%%%%%%%%%%%%%%%%%%%%%%%%%%%%%%%%%%%%%%%%%%%%%%%%
% INTRODUCTION
%%%%%%%%%%%%%%%%%%%%%%%%%%%%%%%%%%%%%%%%%%%%%%%%%%%%%%%%%%%%%%%%%%%%%%%%%%%%%%
Particles of density different from the surrounding fluid 
do not follow the motion of fluid particles, 
and generate inhomogeneous distributions even in incompressible flows
\cite{Re14}.
This phenomenon is crucial in a variety of instances, 
from cloud formation in the atmosphere, to the dynamics of plankton
in the ocean and lakes, to industrial applications e.g. in reactors
\cite{GW13}.
Inhomogeneous distribution in turbulent flows is also of interest 
from a theoretical point of view.  
In recent years, analytical, numerical and experimental studies 
led to significant advances in the understanding of this process
\cite{SE91,BF01,Be03,BD04,BBC07,Fo12}.
In an incompressible turbulent flow, non-inertial, fluid particles follow the 
flow streamlines and remain by definition homogeneously distributed. 
In contrast, inertial particles with density different from the fluid, 
are known to accumulate in regions of high vorticity (light particles)
or high strain (heavy particles) \cite{SE91,Be03,WM93}, as a consequence of 
the accelerations induced by the flow.
Recent analytical and numerical works have shown that gravity interacts
with turbulent accelerations to increase clustering of inertial particles.
Moreover, turbulence can increase the settling velocity 
with respect to still fluid, by pushing particles in regions of downward flow
\cite{BH14,GV14}.
In presence of density fluctuations, gravity also affects the flow itself
as in the case of stratified turbulence which finds many applications in 
natural and technological flows \cite{Li83,RL00}. 
One important example is ocean dynamics which is strongly affected by the 
presence of the pycnocline resulting from temperature and salinity 
variations \cite{Th07,WF11}.

Remarkably, very little is know about the distribution of inertial particles
in stratified turbulence, in spite of its relevance for oceanic and other
applications. Recent works have studied the effect of stratification on the
clustering of heavy \cite{AC08} and light \cite{AC10} particles and 
the effect of a vertical confinement in homogeneous turbulence \cite{De15}.
In this Letter we investigate, by means of direct numerical simulations, 
the distribution of small buoyant particles transported in a turbulent 
stratified flow. We study both the large scale vertical distribution 
and the small scale clustering of particles as a function of the 
relevant parameters. 
Unexpectedly, we find that small scale (fractal) clustering is determined
by the particles relaxation time and is almost independent on the 
degree of stratification. 
Conversely, the vertical confinement of particles is mostly controlled
by the degree of fluid stratification, and weakly depends on the particle
properties. The dependence of the confinement on Froude number is
interpreted within the framework of stratified turbulence phenomenology.

%%%%%%%%%%%%%%%%%%%%%%%%%%%%%%%%%%%%%%%%%%%%%%%%%%%%%%%%%%%%%%%%%%%%%%%%%%%%%%
\begin{figure}[h!]
\begin{center}
\includegraphics[width=1.0\columnwidth]{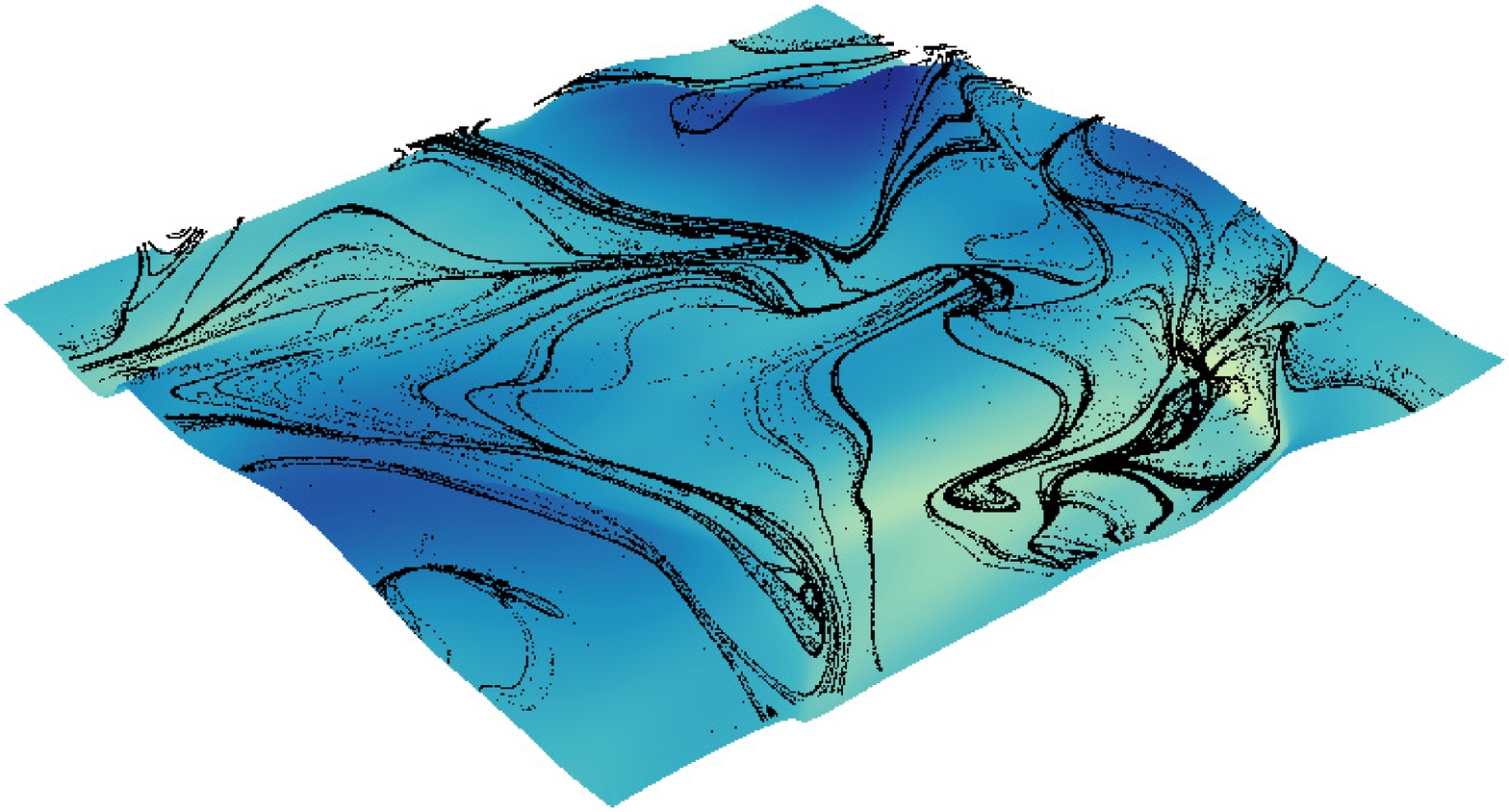}
\vspace{-0.5cm}
\includegraphics[width=1.0\columnwidth]{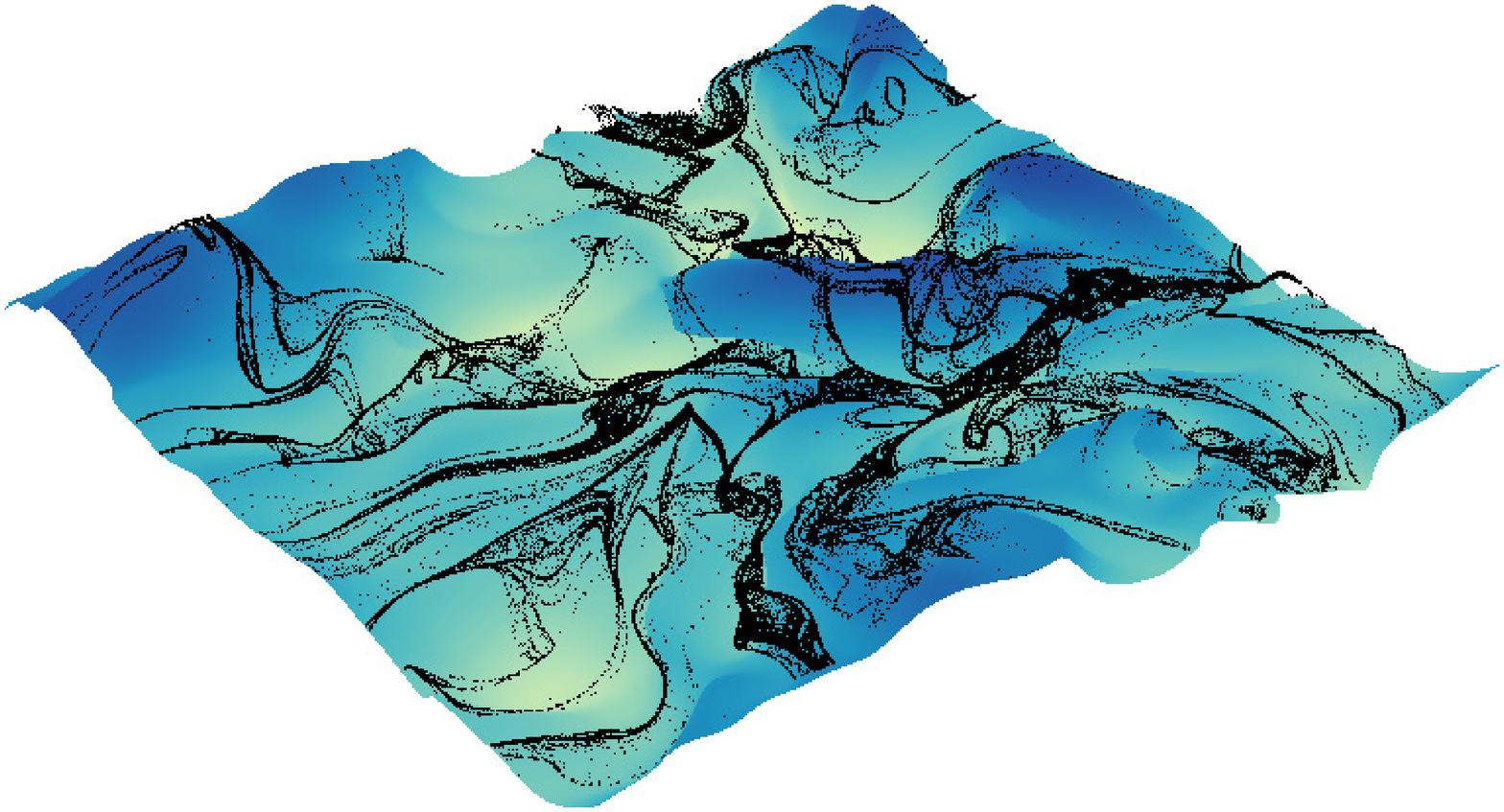}
\caption{Isopycnal surfaces $\theta=z$ corresponding to the particle
density $\rho_0$ for two runs at $Fr=0.2$ (upper plot) and $Fr=0.3$ (lower
plot). The points represent the (fractal) distribution of particles 
with $\tau=0.01$ in the stationary state.}
\end{center}
\label{fig1}
\end{figure}
%%%%%%%%%%%%%%%%%%%%%%%%%%%%%%%%%%%%%%%%%%%%%%%%%%%%%%%%%%%%%%%%%%%%%%%%%%%%%%

%%%%%%%%%%%%%%%%%%%%%%%%%%%%%%%%%%%%%%%%%%%%%%%%%%%%%%%%%%%%%%%%%%%%%%%%%%%%%%
We consider a cubic box of size $L$ of fluid linearly (and stably) 
stratified in the direction $z$ of gravity ${\bf g}=(0,0,-g)$ 
with a constant mean density gradient $d \rho/dz=-\gamma$.
Within the Boussinesq approximation, the motion for the incompressible 
velocity field ${\bf u}({\bf x},t)$ is ruled by
\begin{equation}
\frac{\partial {\bf u}}{\partial t} + {\bf u} \cdot {\bf \nabla} {\bf u} =
- \frac{1}{\rho_0} {\bf \nabla} p + \nu \nabla^2 {\bf u} - 
N^2 \theta \hat{\bf z} + {\bf f}
\label{eq:1}
\end{equation}
\begin{equation}
\frac{\partial \theta}{\partial t} + {\bf u} \cdot {\bf \nabla} \theta =
{\bf u} \cdot \hat{\bf z} + \kappa \nabla^2 \theta
\label{eq:2}
\end{equation}
together with ${\bf \nabla} \cdot {\bf u}=0$. 
The scalar field $\theta({\bf x},t)$, which in the above equations
has the dimension of a length, represents the deviations of the local
density from the linear vertical profile, $\rho=\rho_0-\gamma(z-\theta)$.
$\nu$ is the kinematic viscosity, $\kappa$ the density diffusivity
and $N=(\gamma g/\rho_0)^{1/2}$ is the Brunt-V\"ais\"al\"a frequency.
${\bf f}$ represents an external mechanical forcing needed to sustain 
turbulence.
%In steady conditions ${\bf u}=0$, $\theta=0$ and the density in the 
%box varies between $\rho_0-\gamma L/2$ and $\rho_0+\gamma L/2$. 
In the inviscid, unforced limit ($\nu=\kappa={\bf f}=0$) equations
(\ref{eq:1}-\ref{eq:2}) conserve the total energy, sum of kinetic
and potential contributions, $E=\frac{1}{2} \langle |{\bf u}|^2 \rangle 
+ \frac{1}{2} N^2 \langle \theta^2 \rangle$ where $\langle ... \rangle$ 
denotes the average over the domain.

The velocity ${\bf v}$ of a small inertial particle transported by the flow 
${\bf u}$ generated by (\ref{eq:1}) evolves according to \cite{MR83}
\begin{equation}
\frac{d {\bf v}}{d t} = \beta \frac{d {\bf u}}{d t} - \frac{{\bf v}-{\bf u}}{\tau_p} + (1-\beta) {\bf g}
\label{eq:3}
\end{equation}
where $\beta=3 \rho/(\rho+ 2 \rho_p)$ is the density ratio ($\rho_p$ is the 
density of the particle of radius $a$) and $\tau_p=a^2/(3 \nu \beta)$ 
is the viscous relaxation time. We consider the limit
of small particles (i.e. with small $\tau_p$ for which one can approximate
$\frac{d {\bf u}}{d t} \simeq \frac{d {\bf v}}{d t}$ \cite{MR83})
of density $\rho_p=\rho_0$ so that we can rewrite (\ref{eq:3}) as
\begin{equation}
{\bf v}={\bf u}-\tau_p (1-\beta) \frac{d {\bf v}}{d t}+ \tau_p (1-\beta) {\bf g}
\label{eq:4}
\end{equation}
Consistently with the Boussinesq approximation, in the limit of small
$\tau_p$, we can neglect the second term in the r.h.s. of (\ref{eq:4}) and,
since $(1-\beta){\bf g}\simeq -\frac{2}{3} N^2 (z-\theta)\hat{\bf z}$,
we obtain a simplified expression for the velocity of the floater whose
position ${\bf x}$ evolves according to
\begin{equation}
\frac{d {\bf x}}{dt} = {\bf v}={\bf u}-\frac{1}{\tau} (z-\theta) \hat{\bf z}
\label{eq:5}
\end{equation}
where $\tau \equiv {3/(2 N^2 \tau_p)}$ represents the relaxation
time of the particle onto the isopycnal surface of density $\rho=\rho_0$.
This surface, defined implicitly by the relation $z=\theta(x,y,z)$,
will be denoted as $h(x,y)$, keeping in mind that in general it can
be multivalued.
We remark that more general models for the motion of floaters in stratified 
turbulence are possible, at the price of increased complexity and number 
of parameters \cite{AC10}. 

Although the fluid velocity ${\bf u}$ is incompressible,
the velocity field transporting the floaters is not since 
${\bf \nabla} \cdot {\bf v}=-(1-\partial \theta/\partial z)/\tau$
which is in general nonzero. Formally, this expression represents the 
rate of contraction of the phase space (here the configuration
space) under the dynamics. 
When it is negative we expect that trajectories of floaters
will collapse on a (dynamical) fractal attractor in the phase space.
An example of the attractor is displayed in Fig.~\ref{fig1} which shows
that the {\it large scale} confinement in the vertical direction 
coexists with a {\it small scale} clustering with fractal distribution on 
the isopycnal surface.

%%%%%%%%%%%%%%%%%%%%%%%%%%%%%%%%%%%%%%%%%%%%%%%%%%%%%%%%%%%%%%%%%%%%%%%%%%%%%%

We have integrated the Boussinesq equations (\ref{eq:1}-\ref{eq:2}) 
in a domain of size $L=2 \pi$ with periodic boundary conditions
by means of a fully parallel pseudo-spectral code at resolution up to $N=256$.
Turbulence is generated by a $\delta$-correlated in time isotropic
forcing ${\bf f}$ which is active on a spherical shell of wavenumber 
around $k_f=1$ and which pumps energy at the fixed rate $\varepsilon$.
These parameters define, together with the Brunt-V\"ais\"al\"a frequency, the 
Froude number $Fr \equiv (\varepsilon^{1/3} k_f^{2/3})/N$ which
measures the (inverse) stratification.
We remark that, in stratified turbulence, the mean kinetic energy 
dissipation at small scales $\varepsilon_{\nu}$ is typically 
smaller than the input $\varepsilon$ and depends on $Fr$, because a 
fraction $\varepsilon_{\kappa}$ of the energy input is converted 
into potential energy during the turbulent cascade and dissipated by
diffusivity \cite{SB15}.
Another relevant parameter in stratified turbulence
is the buoyancy Reynolds number
$Re_b=\varepsilon/(\nu N^2)$ defined in terms of the ratio of the buoyancy 
(Ozmidov) scale $\ell_B=\varepsilon^{1/2}/N^{3/2}$ to the dissipative
scale $\ell_D=\varepsilon^{3/4}/\nu^{1/4}$ as $Re_b=(\ell_B/\ell_D)^{4/3}$,
in analogy to the usual Reynolds number $Re=(L/\ell_D)^{4/3}$.
These three numbers are not independent since 
$Re_b = Fr^2 Re$ \cite{BB07} and $Re_b=1$ discriminates between 
stratified-viscous flow (\(Re_b < 1\)) and stratified 
turbulence (\(Re_b > 1\)) \cite{BB07}.
Our simulations are within the turbulent regime, as $Re_b$ is in the 
range $17 \le Re_b \le 540$. 
Together with the (\ref{eq:1}-\ref{eq:2}), we integrated the equation
(\ref{eq:5}) for the particle motion for a set of $10$ classes of particles
characterized by different values of $\tau$ in the range
$0.01 \le \tau \le 10.0$. In presenting the results, this time will be
made dimensionless with the Kolmogorov time 
$\tau_{\eta}=(\nu/\varepsilon)^{1/2}$ by introducing a "Stokes number''
$St \equiv \tau/\tau_{\eta}$.
Table~\ref{table1} reports the most important parameters of the simulations.

%%%%%%%%%%%%%%%%%%%%%%%%%%%%%%%%%%%%%%%%%%%%%%
\begin{table}[t!]
\begin{tabular}{c|c|c|c|c|c|c|c|c}
\hline \hline
$N$ & $\nu, \kappa$ & $k_f$ & $\varepsilon_I$ & $\eta$ & $\tau_{\eta}$ &
$Re$ & $Fr$ & $\tau$ \\ \hline
$128$ & $5 \times 10^{-3}$ & $1.0$ & $0.195$ & $0.028$ & $0.16$ & $430$ & 
$0.2-1.0$ & $0.01-10.0$ \\
$256$ & $4 \times 10^{-3}$ & $1.0$ & $0.195$ & $0.024$ & $0.14$ & $540$ & 
$0.2-1.0$ & $0.01-10.0$ \\
\hline \hline
\end{tabular}
\caption{Parameters of the simulations. $N$ resolution, $\nu$ and $\kappa$ 
kinematic viscosity and diffusivity, $k_f$ forcing wavenumber,
$\varepsilon_I$ energy input rate, $\eta=(\nu^3/\varepsilon_I)^{1/4}$ 
Kolmogorov scale, $\tau_{\eta}=(\nu/\varepsilon_I)^{1/2}$ Kolmogorov
timescale, $Re=U_{rms}L/\nu$ and $Fr=\varepsilon^{1/3} k_f^{2/3}/N$.
$Re$ is computed at $Fr=1$ at which the flow is almost isotropic, 
as $U_{rms}$ varies with $Fr$. The forcing scale is defined as 
$L=\pi/k_f$. All the simulations are performed at 
Schmidt number $Sc=\nu/\kappa=1$.}
\label{table1}
\end{table}
%%%%%%%%%%%%%%%%%%%%%%%%%%%%%%%%%%%%%%%%%%%%%%%%%%%%%%%%%%%%

\begin{figure}[h!]
\begin{center}
\includegraphics[width=1.0\columnwidth]{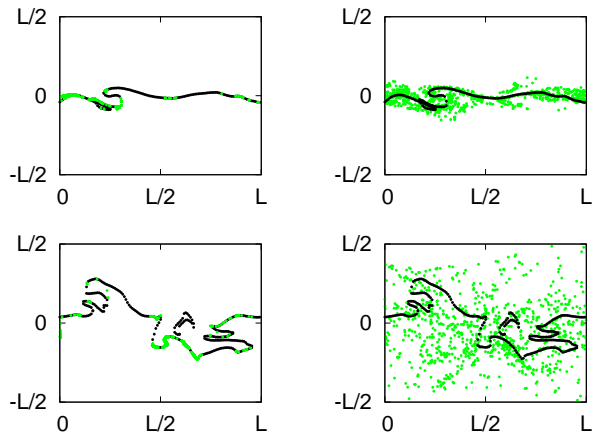}
\end{center}
\caption{(Color online) 
Vertical $(x,z)$ sections of isopycnal surfaces $h$ (black points) for 
$Fr=0.3$ (upper row) and $Fr=1.0$ (lower row)
together with the positions of particles (green points) with $St=0.6$ 
(left column) and $St=60.0$ (right column).
}
\label{fig2}
\end{figure}

In Figure~\ref{fig2} we show vertical sections (at $y=0$) of the isopycnal 
surface $h$ (obtained from the solution of $z=\theta(x,y,z)$) together with 
positions of the particles on the same sections, for different values 
of the parameters $Fr$ and $\tau$. 
It is evident that the isopycnal surface $h$ is almost flat for
strong stratification and it becomes more bent (and multivalued) as $Fr$ 
increases.  Figure~\ref{fig2} shows also the effect of the relaxation time 
$\tau$ on the particles. When $\tau/\tau_{\eta}<1$ the particles are
practically attached to the isopycnal surface, while their 
positions depart from the surface $h$ by increasing $\tau$.

%%%%%%%%%%%%%%%%%%%%%%%%%%%%%%%%%%%%%%%%%%%%%%%%%%%%%%%%%%%%%%%%%%%%%%%%%%%%%%
%%%%%%%%%%%%%%%%%%%%%%%%%%%%%%%%%%%%%%%%%%%%%%%%%%%%%%%%%%%%%%%%%%%%%%%%%%%%%%

%%%%%%%%%%%%%% statistics of the surface

%%%%%%%%%%%%%%%%%%%%%%%%%%%%%%%%%%%%%%%%%%%%%%%%%%%%%%%%%%%%%%%%%%%%%%%%%%%%%%
\begin{figure}[h!]
\includegraphics[width=1.0\columnwidth]{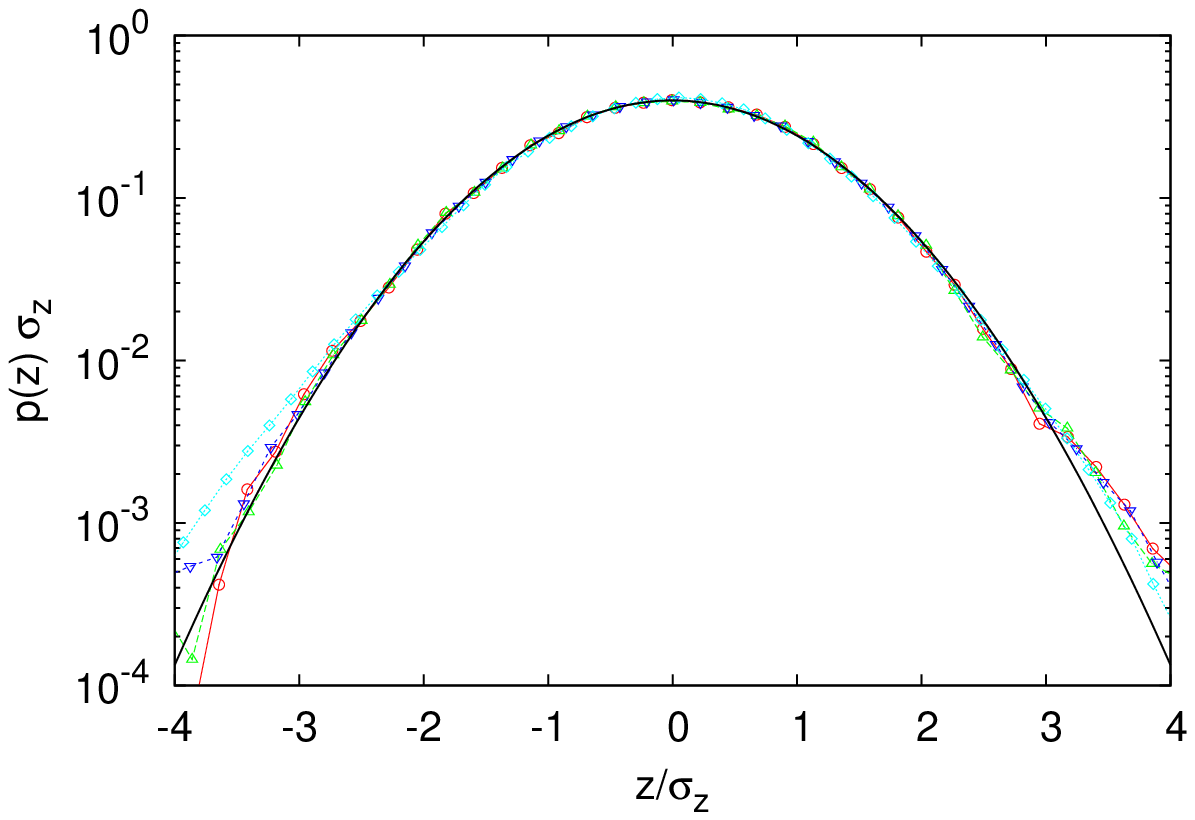}
\includegraphics[width=1.0\columnwidth]{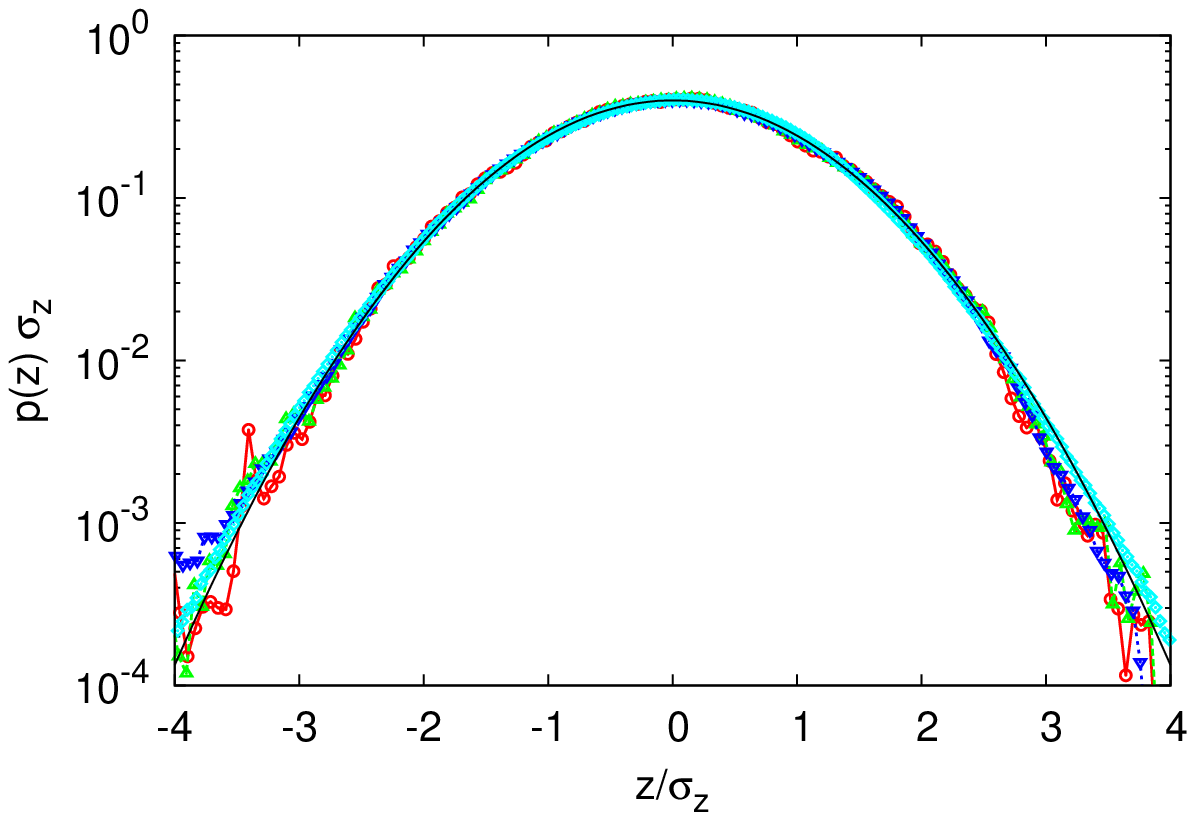}
\caption{Probability density functions of vertical positions of 
tracers for $Re=430$, $Fr=0.3$ (upper plot) and $Re=540$?, $Fr=1.0$
(lower plot). In both cases the PDF for different classes of floaters,
corresponding to $St=0.06$, $0.6$, $6.0$ and $60.0$
are shown.}
\label{fig3}
\end{figure}
%%%%%%%%%%%%%%%%%%%%%%%%%%%%%%%%%%%%%%%%%%%%%%%%%%%%%%%%%%%%%%%%%%%%%%%%%%%%%%

The probability density functions (PDF) of the vertical displacement 
of particles with respect to their equilibrium position $z=0$ in the 
absence of turbulence is shown in Fig.~\ref{fig3} for different values of 
$Re$, $Fr$ and $\tau$. 
We found that in the wide range of parameters investigated these distributions 
are close to Gaussian (with some possible deviations in the tails). 

%%%%%%%%%%%%%%%%%%%%%%%%%%%%%%%%%%%%%%%%%%%%%%%%%%%%%%%%%%%%%%%%%%%%%%%%%%%%%%
\begin{figure}[h!]
\includegraphics[width=1.0\columnwidth]{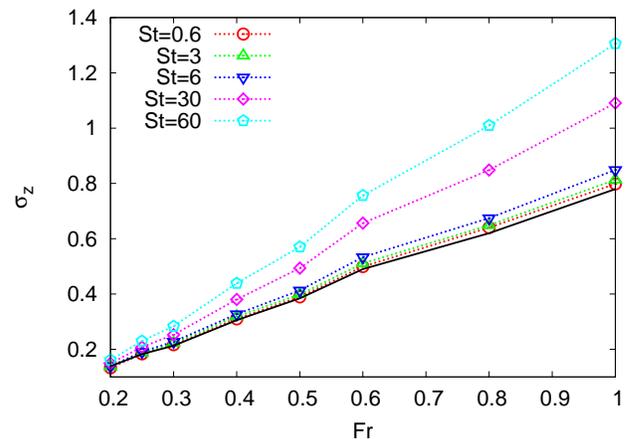}
\caption{Standard deviations $\sigma_z$ of the PDFs of particle 
vertical position as a function of $Fr$ for different values of $St$
at $Re=430$. The solid black line represents the standard deviation
$\sigma_h$ of the isopycnal surface.
}
\label{fig4}
\end{figure}
%%%%%%%%%%%%%%%%%%%%%%%%%%%%%%%%%%%%%%%%%%%%%%%%%%%%%%%%%%%%%%%%%%%%%%%%%%%%%%

The standard deviations $\sigma_z$ of these vertical distribution of particles
for different values of $Fr$ and $St$ are shown in Fig.~\ref{fig4}.
We obtain a linear scaling of $\sigma_z$ on $Fr$, with a coefficient which
shows a (weak) dependency on $St$. For stronger stratification,
$Fr \le 0.3$, the standard deviation is almost independent on $St$, a feature
which can be understood by looking at the plots in Fig.~\ref{fig2}. 
It is indeed evident that for $St$ not too large, the effect of the relaxation
term in (\ref{eq:5}) is to allow the particle to detach from the level 
$z=\theta$ and to remain ``suspended'' for a time of order $\tau$ before
feeling the vertical velocity towards the isopycnal surface.
Because in stratified turbulence vertical velocity and vertical gradient of 
vertical velocity are suppressed \cite{BB07}, there is no mechanism which 
ejects particles far from the surface.
Therefore, as shown by Fig.~\ref{fig2}, the vertical region visited by 
particles reflects the vertical extension of the isopycnal surface and, 
therefore, is independent on $\tau$. 
By increasing $Fr$, the vertical components of the velocity
and of the velocity gradient increase and this produces 
a displacement of particles from the isopycnal surface when the relaxation time
$\tau$ is sufficiently large.

In the limit of small $St$, the standard deviations of particles 
$\sigma_z$ collapse on the standard deviation $\sigma_h$ of the isopycnal 
surface. The linear dependence on $Fr$ shown in Fig.~\ref{fig4}
can be understood within the framework of stratified turbulence, as a 
manifestation of the presence of the so-called {\it vertical shear layers}
\cite{BC01} and the associated vertical correlation scale of velocity $L_v$. 
Physically this scale represents the vertical displacement for converting 
injected kinetic energy into potential energy and can be estimated simply 
as $L_v \simeq U/N$ ($U$ is a typical large scale velocity) and therefore 
one obtains $L_v \propto Fr$, i.e. linear scaling as shown in Fig.~\ref{fig4}.

%%%%%%%%%%%%%%%%%%%%%%%%%%%%%%%%%%%%%%%%%%%%%%%%%%%%%%%%%%%%%%%%%%%%%%%%%%%%%%
\begin{figure}[h!]
\includegraphics[width=1.0\columnwidth]{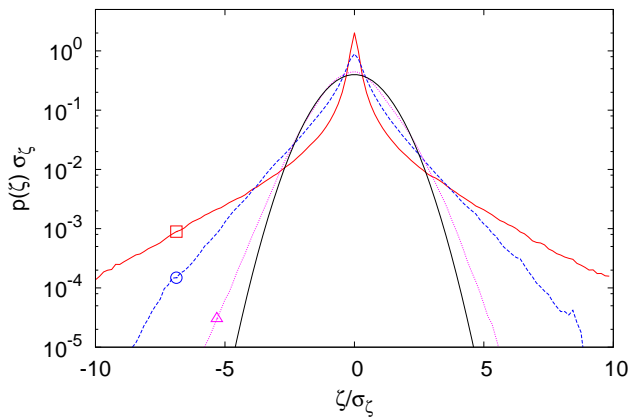}
\caption{Probability density functions of the variable $\zeta=z-\theta$
for $Re=540$, $Fr=1.0$ and $St=1.4$ (red line with square), 
$St=7.0$ (blue line with circle) and $St=70$ (pink line with triangle).
The black line is a Gaussian.}
\label{fig5}
\end{figure}
%%%%%%%%%%%%%%%%%%%%%%%%%%%%%%%%%%%%%%%%%%%%%%%%%%%%%%%%%%%%%%%%%%%%%%%%%%%%%%

The deviation of the particles form the isopycnal surface can be 
investigated by looking at the statistics of the variable 
$\zeta=z-\theta$. The evolution of this quantity is obtained by 
the time evolution of the field $\theta$ along the trajectory 
of a floater moving with the velocity ${\bf v}$ and which gives 
(neglecting the diffusive term)
$d(z - \theta)/dt = -(z-\theta) (1-\partial \theta/\partial z)/\tau$.
In the absence of fluctuations ($\theta=0$) this equation would simply
represent the linear relaxation of particles towards the isopycnal layer
$z=0$ \cite{De15}.
This is not achieved since the term $\partial \theta/\partial z$ 
is fluctuating without a definite sign.
Figure~\ref{fig5} shows the normalized PDF of the variable $\zeta$ 
for three different values of $\tau$ at $Fr=1.0$. It is evident that
the statistics is neither Gaussian nor scale invariant and the PDF develops 
large tails for small relaxation times.
 %%%% Wile E. Coyote effect %%%%%
These large fluctuations are due to the folding of the
isopycnal surface. 
%Particles located in the neighborhood of a fold in which the
%isosurface is almost vertical (see Figure~\ref{fig2})
%can suddenly be at large distance in the vertical direction from the
%nearest branch of the surface below or above them, 
%in spite of being still very close in the horizontal direction 
%to the branch of the surface that they were belonging to. 
Particles located in the neighborhood of a fold in which the
isosurface is almost vertical (see examples in Figure~\ref{fig2}) 
are not restored horizontally to the close-by branch they just left, 
but rather displaced vertically by buoyancy 
toward the nearest branch above or below them. 
This mechanism, which is enhanced at large $Fr$, 
produces a sudden increase 
in the distance between the particles and the isopycnal surface, 
and causes the development of large tails in the PDF of $\zeta$.  
%%%%%%%%%%%%%% fractal dimension

%%%%%%%%%%%%%%%%%%%%%%%%%%%%%%%%%%%%%%%%%%%%%%%%%%%%%%%%%%%%%%%%%%%%%%%%%%%%%%
\begin{figure}[h!]
\includegraphics[width=1.0\columnwidth]{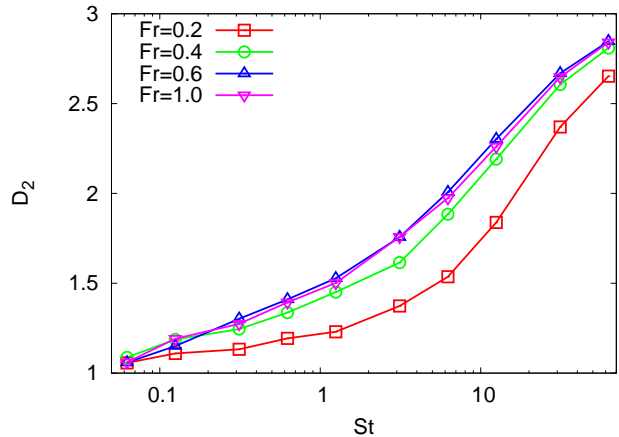}
\caption{Correlation dimension $D_2$ versus relaxation time $St$ for different 
values of stratification parameter $Fr$.}
\label{fig6}
\end{figure}
%%%%%%%%%%%%%%%%%%%%%%%%%%%%%%%%%%%%%%%%%%%%%%%%%%%%%%%%%%%%%%%%%%%%%%%%%%%%%%

As already discussed, floating particles moving according to (\ref{eq:5})
are transported by a compressible velocity field and are therefore expected to
relax on a (dynamical) fractal subset of the physical space, as shown in the
examples of Fig.~\ref{fig1}. 
In order to characterize this subset, and its dependence
on the parameters, we have measured the correlation dimension $D_2$ of particle
distribution, defined as the scaling exponent of the probability of finding two
particles at distance less than $r$: $P(|{\bf x}_1-{\bf x}_2|<r) \propto r^{D_2}$
as $r \to 0$ \cite{PV87}. 
The maximum value $D_2=3$ denotes uniformly distributed
particles, while $D_2<3$ indicates fractal patchiness with smaller 
$D_2$ corresponding to more clustered distributions and increased 
probability of finding pairs of particles at close separation.
Figure~\ref{fig6} shows $D_2$ as a function of $St$ for different values 
of $Fr$. 
In the limit $St \gg 1$, for which ${\bf v} \to {\bf u}$ in (\ref{eq:5}), 
floaters move as fluid particles in an incompressible velocity and 
therefore remain uniformly distributed in the volume with $D_2=3$. 
The fractal dimension is found to be monotonic in $St$ and attains a 
minimum value $D_2 \simeq 1$ for the smallest relaxation time. 
This value indicates distributions of particles on quasi-one-dimensional
structures (as shown qualitatively in Fig.~\ref{fig1}) which is almost
independent on $Fr$. This is a remarkable result, as one could expect that 
for strong stratification and small $\tau$, in which particles lie on 
the isopycnal surface which is almost flat, the fractal dimension would be 
close to $D_2=2$. The fact that we find $D_2 \simeq 1$ indicates that 
the dynamics on the surface is dissipative and therefore the horizontal 
velocity field is compressible.

The weak dependence of the correlation dimension on the stratification 
extends also to larger values of $St$ and we find that, in general, 
$D_2$ is virtually independent on $Fr$ for $Fr \ge 0.5$. This is in contrast 
with the (large scale) vertical confinement shown in Fig.~\ref{fig5} 
which displays a strong (linear) dependence on $Fr$ 
and a weak dependence on $St$.

%%%%%%%%%%%%%% conclusions
In conclusion, we have investigated the dynamics of neutrally buoyant 
particles in a turbulent stratified flow.
We have shown that the extension $\sigma_z$ of the vertical layer in which the 
particles are confined depends on the characteristics of the flow and 
moderately on the size of the particles.
On the contrary, the small scale patchiness inside this layer, expressed 
by the correlation dimension, strongly depends on the particle size and 
only weakly on the stratification of the flow.

One of the most remarkable examples of confinement of particles in the
ocean is the formation of the so-called thin phytoplankton layers (TPL):
aggregations of phytoplankton and zooplankton at high concentration with 
thickness from centimeters to few meters, extending up to several 
kilometers horizontally and with timescale from hours to days \cite{DS12}.

Among the different mechanisms proposed for the formation of TPLs,
buoyancy force in stratified flow is particularly relevant for non-swimming
species and aggregates, such as diatoms and marine snow, which are often 
observed to accumulate in correspondence of strong stratification \cite{Al02}.
In order to test the applicability of our model in the aquatic ecosystems 
we provide a simple example using the typical values observed 
for diatom-dominated marine snow \cite{Al02}.
Assuming an energy dissipation rate $\varepsilon \sim 10^{-8}~{\rm m}^2 {\rm s}^{-3}$
($\tau_{\eta} \simeq 10~{\rm s}$) and a Brunt-V\"ais\"al\"a frequency 
$N \sim 0.1~{\rm s}^{-1}$, aggregates of size $a \simeq 0.5 \, {\rm cm}$ have a 
relaxation time $\tau \simeq 18~{\rm s}$ which corresponds to $St=1.8$,
inside the range in which we observe clustering at small scales 
(Fig.~\ref{fig6}) for all the values of $Fr$. Smaller aggregates or 
single cells of size $a=0.1 \, {\rm cm}$ correspond to $St \simeq 45$ and,
according to our results, would distribute almost homogeneously 
(with a fractal dimension close to $3$) within the thin layer.

Beside the applications to thin layers, our results are of general interest
as they describe quantitatively and for the first time how the combination 
of turbulence and stratification generates both large scale confinement and
small scale fractal patchiness in a suspension of buoyant particles. 

We acknowledge the European COST Action MP1305 ``Flowing Matter''.

%%%%%%%%%%%%%%%%%%%%%%%%%%%%%%%%%%%%%%%%%%%%%%%%%%%%%%%%%%%%%%%%%%%%%%%%%%%%%%

%%%%%%%%%%%%%%%%%%%%%%%%%%%%%%%%%%%%%%%%%%%%%%%%%%%%%%%%%%%%%%%%%%%%%%%%%%%%%%

\end{document}